# On the thermodynamics of DNA double strand in the Peyrard-Bishop-Dauxois model

Likhachev I.V., Shigaev A.S., Lakhno V.D.

*The Institute of Mathematical Problems of Biology, Russian Academy of Sciences – a branch of the KIAM RAS. Address: IMPB RAS, 1, Professor Vitkevich St., 142290, Pushchino, Moscow Region, Russia*

**Abstract**

The temperature dependence of energy, partition function, entropy and free energy, describing a first-order-like phase transition of a DNA molecule is calculated based on the direct method of molecular dynamics for the classical Peyrard-Bishop-Dauxois model. It is shown that without taking into account the quantum freezing of the degrees of freedom at low temperatures, the relationship between these quantities is not determined. The latent heat of DNA melting is calculated for homogeneous PolyA/PolyT and PolyG/PolyC DNA chains. It is shown that this melting transition corresponds to a sharp temperature dependence of the heat capacity.

## 1. Introduction

Currently, one of the most important mechanical models used to study the equilibrium and nonequilibrium properties of DNA is the Peyrard-Bishop-Dauxois (PBD) model [1–7]. Despite its simplicity, this model properly describes the sharp temperature dependence of melting of long homogeneous DNA fragments, multi-stage melting of heterogeneous DNA fragments [8,9], and also provides a reasonable relative statistics of the formation of denaturation bubbles in double-stranded DNA [10]. These properties of the DNA molecule play an important role in the processes such as transcription, denaturation, replication and recombination [11,12].

The knowledge of the thermodynamic properties of DNA is also important for biotechnology applications, such as the design of probes and targets for primers, polymerase chain reactions, and molecular beacons. In addition, the knowledge of thermodynamic characteristics is important in nanotechnology and nanobioelectronics, in the design of nanoelectronic devices and creation of molecular memory [13–16].

The aim of this work is to calculate thermodynamic characteristics using direct simulations by molecular dynamics (MD) method which was used in [17,18] to calculate the heat capacity of DNA based on the PBD model. To do this, when calculating such characteristics as the partition function, entropy and free energy based on the classical PBD model, a cutoff is introduced which takes into account the freezing out of classical degrees of freedom at low temperatures (See Suppl.A).

Below we present the results of calculating the main thermodynamic quantities of DNA for the parameters of the PBD model used in [2,3], which are most adequate to the currently available experimental data.

## 2. Model and parameters

Hamiltonian of the PBD model has the form [1–7]:

$$H = \sum_{n=1}^{N} \frac{M}{2} \dot{y}_n^2 + \sum_{n=1}^{N} \left\{ D_n \left( e^{-\alpha_n y_n} - 1 \right)^2 \right\} + \sum_{n=1}^{N-1} \left\{ \frac{k}{2} \left( 1 + \rho e^{-\beta(y_n + y_{n+1})} \right) \left( y_n - y_{n+1} \right)^2 \right\}, \quad (1)$$

where $M$ is the effective mass of a nucleotide pair, $D_n$ and $\alpha_n$ – the depth and inverse width of the potential well for interactions in a complementary nucleotide pair, respectively. The elastic constant $k$ takes into account stacking interactions, $\beta$ is the decay constant of stacking interactions, while $\rho$ is a dimensionless parameter that phenomenologically takes into account the denaturation cooperativity of a π-stack [1]. Summation is performed over all $N$ nucleotide pairs, in the PBD model, $y_n$ corresponds to the relative displacement of nucleotides from their equilibrium positions in the $n$-th nucleotide pair. According to the specifics of the model, $y_n$ corresponds to the real elongation of H-bonds in a complementary base pair multiplied by √2, see [19].

The first term in (1) corresponds to the kinetic energy of oscillations of nucleotide pairs, the second and third terms correspond to the potential energies of on-site and inter-site interactions. The second term of formula (1) is the Morse potential. It describes the hydrogen bonds between nucleotides in a pair. The third term corresponds to the stacking interaction of adjacent nucleotide pairs.

We choose the common effective mass as $M = 300.5$ a.m.u.; other parameters were taken from the paper by Campa and Giansanti: $\beta = 0.35$ Å$^{-1}$, $k = 0.025$ eV/Å$^2$, $\rho = 2.0$, $D_{AT} = 0.05$ eV, $\alpha_{AT} = 4.2$ Å$^{-1}$, $D_{GC} = 0.075$ eV, $\alpha_{GC} = 6.9$ Å$^{-1}$ [2].

## 3. Methods and results

To simulate a DNA chain of an arbitrary primary structure, the molecular dynamics method was used. The integration of the equations of motion was carried out according to the Verlet integration [20]. To maintain the temperature in the system, a collision thermostat was used [21–23] (See Suppl B).

In [17,18], we calculated the melting temperature of DNA based on the PBD model (1). For this purpose, in [17,18], the temperature dependence of the chain energy $E(T)$ was calculated and the temperature dependence of its heat capacity was found.

The results obtained in [17,18] make it possible to construct the entire thermodynamics for the PBD model. Knowing $E(T)$, we can calculate the free energy of the chain F using the known thermodynamic relation:

$$E = -T^2 \left( \frac{\partial}{\partial T} \cdot \frac{F}{T} \right). \quad (2)$$

From (2) follows the expression for the free energy F:

$$\frac{F}{T} = -\int_{\omega}^{T} \frac{E(\tau)}{\tau^2} d\tau \quad (3)$$

where:

$$\omega = \sqrt{\frac{2D\alpha^2}{M}}, \quad (4)$$

ω is the frequency of oscillations of a nucleotide pair at the bottom of the potential well. Here we put $k_B = 1$, $\hbar = h/2\pi = 1$, were $k_B$ is the Boltzmann constant and h is the Planck constant.

Limitation of the lower limit in (3) by the frequency of harmonic oscillations ω takes into account the quantum mechanical freezing of vibrational degrees of freedom at $T=\omega$. Consideration of quantum freezing removes the divergence at the lower temperature limit in integral (3), allowing us to reveal a relation between the main thermodynamic characteristics of the Hamiltonian of the PBD model (Suppl.A).

Taking account of the above parameter values, we obtain for ω in (4): ω=57.5 K for AT-pairs and ω=115.8 K for GC pairs. To calculate integral (3), it is necessary to find the energy per each nucleotide pair (site). To do this, during the MD experiment, the energy matrix $E_{n,t}$ is recorded as an output value for each realization, where n is the site number, $t$ is the time instant, with subsequent averaging of $E_{n,t}$ over the realizations. When integrating the motion equations for the energy matrix, $E_{n,t}= 0$ was assumed for a base pair if the temperature of the system is less than the relevant lower limit of integral (3). Such accounting, however, has virtually no effect on the calculations of the melting temperature of DNA and the temperature dependence of its heat capacity, performed in [17,18], since such a calculation refers to a temperature which significantly exceeds the lower limit of integration in (3).

Figures 1, 2 show the dependences $f(T) = F(T)/N$, were $f$ is the free energy per a nucleotide pair for $E(T)$ found by the MD method described in [17,18] and related to $F(T)$ by relation (3), $\varepsilon(T) = E(T)/N$ and entropy $s(T)$, determined by $s=S/N$. Thus, the quantity f has the meaning of the chemical potential of the system μ.

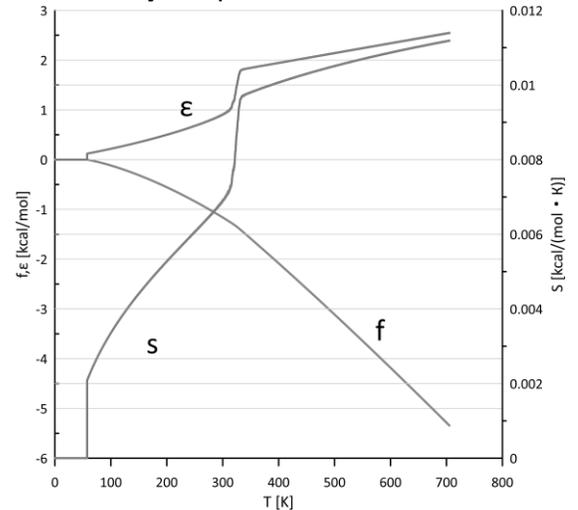

Fig. 1. Temperature dependence of the free energy $f$, energy ε and entropy $s$ per a Watson-Crick pair for an AT chain of 100 base pairs.

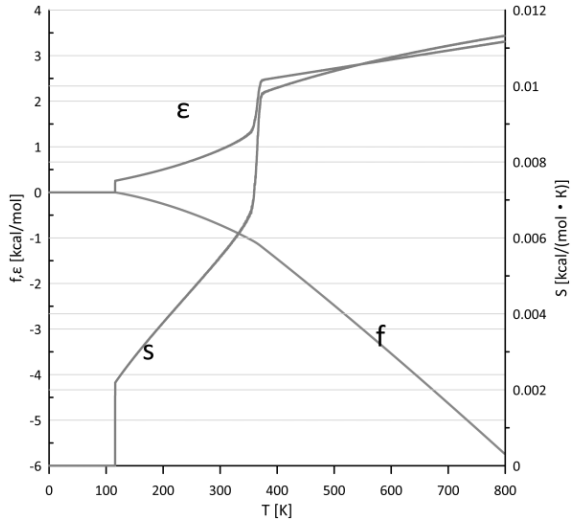

Fig. 2. Temperature dependence of the free energy $f$, energy $\varepsilon$ and entropy $s$ per a Watson-Crick pair for a GC chain of 100 base pairs.

The temperature range in Figure 1 near 325 K corresponds to melting of a chain of AT pairs and that in Figure 2 near 371 K corresponds to melting of a chain of GC pairs. An important feature of melting of a DNA molecule is that it does not occur at one fixed point, but is diffused over a certain temperature range (see, for example, [24]).

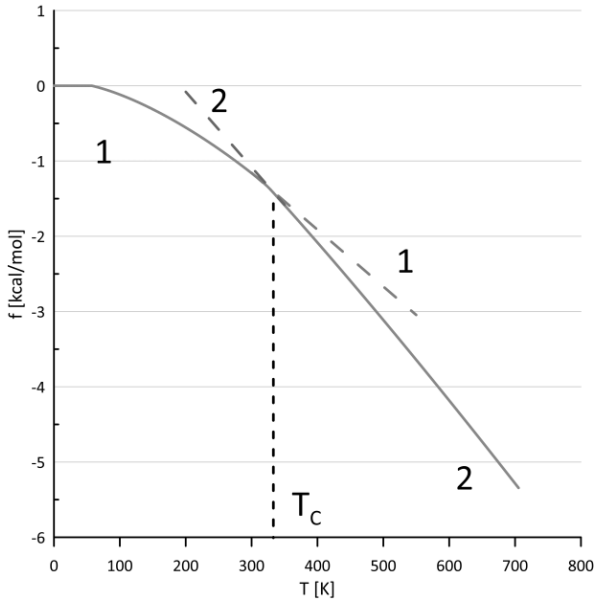

Fig. 3. Temperature dependence of the free energy $f$ per a Watson-Crick pair for AT chain of 100 base pairs near critical point.

Figure 3 shows the free energy curves f(T) near the DNA melting temperature Tc.

Figure 3 suggests that at T<Tc the double-stranded DNA is stable (curve 1), while at T>Tc the denatured DNA is stable (curve 2).

At the point of intersection of the curves at $T=T_C$, the value of the derivative $\frac{\partial f_1}{\partial T} = \frac{\partial \mu_1}{\partial T}$ is greater than the value of $\frac{\partial f_2}{\partial T} = \frac{\partial \mu_2}{\partial T}$. This discontinuity of the derivative at the point $T_c$ means that we are dealing with a first-order-like phase transition, when during the transition from a bound double-stranded state to a denatured one, heat absorption occurs, illustrated by the dependence $\varepsilon(T)$ near $T_c$. Notice that the energy-temperature dependence which has been earlier presented in [25] does not show the phase transition discussed in this work.

The latent heat of this transition $Q$ is determined by the relation: $Q = T_c \Delta S$, where $\Delta S$ is a change in the entropy at $T=T_c$. For an AT pair, this value, according to Figure 1, is $Q$=0.78 kcal/mol. Accordingly, for the GC pair $Q$=0.98 kcal/mol. The results obtained are in agreement with the data available [26].

Note that above $T_c$, curve 1 (the dotted part of this line) corresponds to the metastable state of the double-stranded DNA. Below $T_c$, the metastable state corresponds to curve 2, which represents the denatured state (dotted part of line 2). This follows from the fact that this curve coincides with that if in the expression for the Hamiltonian (1) all $y_n$ tend to infinity, that is all hydrogen bonds between the DNA chains are considered to be broken. At temperatures above the temperature of this stacking interaction, the temperature dependence of the free energy corresponds to the expression:

$$f(T) = -T \ln \left(\frac{T}{\omega}\right),$$

that is, the limiting case of independent nucleotide particles.

The dependence $f(T)$ obtained allows us to calculate the statistical sum $z$ of one nucleotide pair of the AT chain from the ratio:
$$f = -T \ln z \qquad (5)$$
With the use of the thermodynamic expression F=E-TS, the entropy of the AT and GC chains will be:
$$s=(\varepsilon-f)/T, \qquad (6)$$
where $\varepsilon$ and $s$ are related to the total energy of the chain $E$ and entropy $S$ by the relations $\varepsilon=E/N$, $s=S/N$ which temperature dependence is visualized by Fig.1,2.

In [17,18], we calculated the heat capacity of DNA for the parameters taken from paper [1]. Figure 4 shows the temperature dependence of the heat capacity $C_v = dE/dT$ calculated for the parameters from the paper by Campa [2] which are used in this paper.

The temperature dependence of the heat capacity determines such nonequilibrium characteristics as fluctuations of thermodynamic quantities of DNA, their distribution densities, correlation functions that can be experimentally measured from neutron or X-ray scattering, as well as kinetic coefficients.

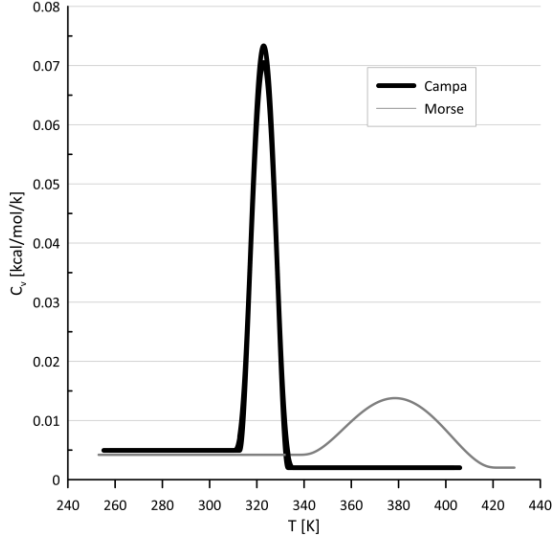

Fig. 4 Temperature dependence of the heat capacity per a Watson-Crick pair for an AT chain of 100 base pairs. Comparison between Morze [1,17,18] ($D_{AT}$=0.04 eV, $\alpha_{AT}$=4.45 Å$^{-1}$, $\beta$=0.35 Å$^{-1}$, $\rho$=0.5, $k$=0.06 eV/Å$^2$) and Campa [2,3] ($D_{AT}$=0.05 eV, $\alpha_{AT}$=4.2 Å$^{-1}$, $\beta$=0.35 Å$^{-1}$, $\rho$=2.0, $k$=0.025 eV/Å$^2$) parameters.

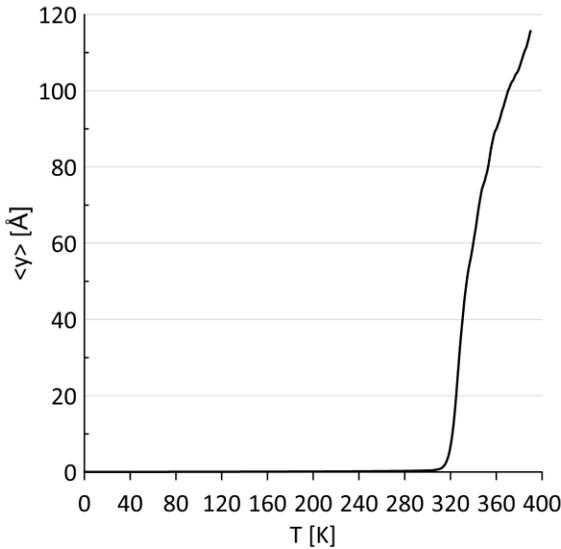

Fig. 5 Temperature dependence of the mean value of the coordinate for an AT chain of 100 base pairs near critical point.

The results obtained by our method are given in Fig. 5 which illustrates the temperature dependence of the mean coordinate value y(T) of a nucleotide pair displacement from the equilibrium position over the realizations for parameters of works [2,3]. In [2,3], the dependence $y(T)$ was used to calculate the average fractions of open and closed DNA states, followed by selection of the PBD model parameters to compare the results of DNA melting calculations with experiment.

It is important to emphasize that, although the solid part of curve 2 for $T$ above $T_c$ describes the denatured state of the DNA molecule, the stacking interaction continues to play a role in it. It follows from Figure 5, that due to the stacking interactions the displacements remain finite even at temperatures higher than melting.

## 4. Discussion

Though the thermal DNA melting is an old problem, the order of its phase transition is not fully understood. There are both suggestions that this is the first-order transition [19,27–29], and the second-order transition [30,31].

Traditionally, statistical thermodynamics of the PBD model of DNA is presented on the basis of the transfer integral method [32–34]. This approach allows us to obtain an analytical solution for a homogeneous chain in the thermodynamic limit, when $N$ tends to infinity. In the case of finite and heterogeneous chains, this approach is poorly suitable. Since the description by the transfer integral method is associated with finding only the ground state in the thermodynamic limit, it cannot be used to describe excitations such as bubbles. The approach we used to construct the thermodynamics for the PBD model is free from these limitations.

Figures 1, 2 suggest that the (diffused) break points in the temperature dependences $\varepsilon$ $(T)$, $f(T)$, $s(T)$ correspond to the melting temperatures of the AT and GC chains. This behavior is typical for the first-order phase transitions.

Notice that an extended model of PBD which takes into account the helicoidal structure of DNA was considered in [35–37]. This model also shows the first order phase transition of DNA denaturation [38].

According to the classical theory, anomalies of thermodynamic quantities should appear at one strictly defined temperature, while, according to the results obtained, they extend over a finite range of changes in various parameters (temperature, ion concentration, PH, external field, etc.). In this regard, the term "first-order phase transition" that we use differs from that used by classical thermodynamics where, according to Ehrenfest [39,40], the condition for a phase transition is a discontinuity of some derivative of a relevant thermodynamic potential. The reason for this difference is thermal fluctuations which lead to blurring of the phase transition and, in particular, to the finite value of the heat capacity in the vicinity of the phase transition (Fig. 4). For the same reason, in relation to DNA, the Landau-Lifshitz theorem on the impossibility of phase transitions in one-dimensional systems is not applicable, since it applies only to point phase transitions according to the Ehrenfest classification.

To summarize, it should be noted that the dependences found for energy, free energy and entropy in the classical PBD model obtained with quantum freezing describe a diffused first-order phase transition from the double-stranded to the denatured state of DNA. This diffuseness is due to the finite heating rate of the thermostat during temperature modeling (as a consequence, the existence of hysteresis and the presence of bubbles) and the limited number of realizations used to calculate the thermodynamic characteristics, as well as the problems associated with Monte Carlo and molecular dynamics simulations near the phase transition point [41] (Suppl.B).

**Acknowledgements**

The authors express their gratitude to N.K. Balabaev for his proposals on the technique of carrying out the numerical experiments.

The research was carried out using the hybrid supercomputer K-60 at the Keldysh Institute of Applied Mathematics, Russian Academy of Sciences.